\documentclass[10pt,reqno]{amsart}

\def\showfigures{1}
\def\showtables{1}

\usepackage{amsmath,amssymb,amsfonts,dsfont}
  \allowdisplaybreaks
\usepackage{cite,graphicx,xcolor,hyperref} 

\theoremstyle{plain}

\theoremstyle{definition}

\theoremstyle{remark}
  \newtheorem{remark}{Remark}

\newcommand{\LPT}{{L_2(\mathds{T})}}
\newcommand{\LPTs}{{L_2(\mathds{T}^*)}}

\newcommand{\BasisT}{\{q(i,t)\}_{i=0}^\infty}

\newcommand{\trans}{{\scriptscriptstyle \mathrm{T}}}

\newcommand{\mi}{\mathrm{i}}
\renewcommand{\Re}{\mathop{\mathrm{Re}}}
\renewcommand{\Im}{\mathop{\mathrm{Im}}}

\newcommand{\bs}{{\scriptscriptstyle \mathrm{BS}}}
\newcommand{\pp}{{\scriptscriptstyle \mathrm{P}}}
\newcommand{\ld}{{\scriptscriptstyle \mathrm{L}}}

\newcommand{\bm}{{\scriptscriptstyle \mathrm{BM}}}
\newcommand{\ha}{{\scriptscriptstyle \mathrm{H}}}
\newcommand{\us}{{\scriptscriptstyle \mathrm{U}}}

\author{K.\,A. Rybakov}

\title[Applying the Spectral Method for Modeling Linear Filters]{Applying the Spectral Method\\for Modeling Linear Filters:\\Bessel, Papoulis, and Legendre Filters}

\begin{document}

\maketitle

\begin{center}

\vskip -3.5ex

Moscow Aviation Institute (National Research University);\\
125993, Moscow, Volokolamskoe Hwy, 4;\\
rkoffice@mail.ru
\end{center}

\vskip 2.5ex

\textbf{Abstract.} This paper proposes a new technique for computer simulation of linear filters. It allows simulating continuous-time linear filters based on the spectral method for analyzing linear control systems. Applying the spectral method implies that the input and output signals are represented by ordered sets of expansion coefficients in a chosen basis, and the filter itself is specified by a two-dimensional nonstationary transfer function. The described technique is tested on Bessel, Papoulis, and Legendre filters of various orders. For each filter, the corresponding two-dimensional nonstationary transfer function, i.e., the matrix of the linear transformation relating expansion coefficients of the input and output signals, is obtained.

\vskip 0.5ex

\textbf{Keywords:} signal processing, spectral method, spectral form of mathematical description, Bessel filter, Papoulis filter, Legendre filter

\vskip 0.5ex

\textbf{MSC:} 42C10; 94A12

\section{Introduction}\label{secIntro}

We consider the problem of signal filtering by frequency division, i.e., the selective transmission of signals in a given frequency range. In this context, a filter is understood as a linear dynamic system that can be described by a linear differential equation with constant coefficients. The purpose of filtering is to extract the useful signal from a mixture of the useful signal and noise. For such filters, no optimality criterion in the state space is specified, as in optimal filtering theory \cite{BaiCri_09, KudRyb_Springer21, VasIsaSteTop_JCSSI25}; instead, the filter quality depends on the properties of the corresponding amplitude-phase frequency response~\cite{Lam_82, Paa_03}.

The main objective of this work is to develop a technique for modeling filters (hereafter referred to as linear filters) that does not require a transition to discrete time. This technique is based on the spectral form of mathematical description of linear control systems \cite{SolSemPeshNed_79}. Previously, it was applied to Butterworth, Linkwitz--Riley, and Chebyshev filters \cite{RybShe_FraOp26}. In this paper, the scope of the spectral form of mathematical description is extended to linear filters defined through Bessel and Legendre polynomials.

Classical orthogonal polynomials are used in various branches of mathematics and its applications. They are successfully employed as basis systems for representing functions in spectral methods \cite{Boyd_01, CanHusQuaZan_06}. Orthogonal polynomials are solutions of second-order linear homogeneous differential equations of a certain type. We can distinguish Jacobi polynomials and their particular cases: Legendre and Chebyshev polynomials (orthogonal on the interval $[-1,1]$), Laguerre polynomials (orthogonal on the half-line $[0,+\infty)$), and Hermite polynomials (orthogonal on the real line $\mathds{R} = (-\infty,+\infty)$). In particular, Chebyshev polynomials of the first kind underlie the commonly used Chebyshev filters of types I and II, while Legendre polynomials and related functions define optimal ``L''-filters developed by Athanasios Papoulis \cite{Pap_PIRE58, Pap_PIRE59} and Legendre filters proposed by Yu Hsiu Ku and Meir Drubin \cite{KuDru_JFI62}. Optimal ``L''-filters are called Papoulis filters or Legendre--Papoulis filters.

Bessel polynomials also belong to the class of orthogonal polynomials, but their orthogonality is understood in the sense of integration over the unit circle in the complex plane $\mathds{C}$. Their orthogonality (with a weight) on the half-line $[0,+\infty)$ is discussed in \cite{EvaEveKwoLit_JCAM93}. Compared to orthogonal polynomials listed above, they are used less frequently. Nevertheless, Bessel polynomials are successfully applied in signal filtering; namely, they define Bessel filters, also called Thomson or Bessel--Thomson filters. In the paper \cite{Tho_PIEE49}, Wilfrid Ernest Thomson used Lommel polynomials, but nowadays it is more convenient to use Bessel polynomials \cite{KraFri_TAMS49} for the mathematical description of such filters.

Thus, there are several families of filters based on orthogonal polynomials and their properties (filters in each family differ by order). For their computer simulation, a transition to discrete time is usually employed~\cite{LutTosEva_01, OppSch_14, Gir_17}. This approach has drawbacks, such as frequency aliasing (for the impulse invariance method) or frequency warping (when using the bilinear transformation). They are described in detail in~\cite{OppSch_14}. The use of the spectral form of mathematical description allows modeling linear filters in continuous time, which can be useful at the filter design stage.

In the context of the spectral method, orthogonal polynomials are of considerable interest. They can be used as a basis for representing the input and output signals and for describing linear dynamic systems in the spectral domain. For the analysis of linear control systems over finite time intervals, Legendre and Chebyshev polynomials are used \cite{SolSemPeshNed_79}; however, there are also general results obtained for Jacobi polynomials and their special case, Gegenbauer polynomials (Legendre and Chebyshev polynomials are themselves special cases of Gegenbauer polynomials) \cite{Rybin_AMI83}.

In this work, orthogonal polynomials are used only for the mathematical description of linear filters and are not used as a basis. However, the choice of basis is important at the stage of the numerical experiment, while the description of filters in the spectral domain is general and independent of the basis. For the numerical experiment in this paper, the trigonometric basis is chosen, but there are no restrictions on the use of orthogonal polynomials.

The presented results complement the paper \cite{RybShe_FraOp26}, which provides a brief overview of modern applications of linear filter theory using Butterworth and Chebyshev filters as examples (the early stages of this theory are described in detail in reviews~\cite{Bel_PIRE62, Dar_TCS84}). In \cite{RybShe_FraOp26}, Linkwitz--Riley filters are additionally considered, but they are directly related to Butterworth filters.

Bessel filters are used in audio signal processing~\cite{ZhaChenWang_IEEE10, CecBruNobTerVal_JAES23}, but they can also be applied in control theory \cite{WangWang_AMM13} and telecommunications \cite{JiaHeWu_IEEE01}. The application area of Papoulis filters includes systems where a steep roll-off of the amplitude frequency response without oscillations (such as those characteristic of Chebyshev filters) is important. In particular, these are modern 5G/6G communication systems \cite{SanCra_RE25}. In addition, image processing methods based on Papoulis filters are used in biological and medical research \cite{ArbArbDubZolLapLuk_OIP23, Xu_SPIE26}. Legendre filters find applications in control system design \cite{SamDenTit_SFedU24}.

Note that Butterworth and Chebyshev filters are used much more often than Papoulis and Legendre filters in various applications. Papoulis and Legendre filters have better properties; however, their synthesis involves a more complex derivation of transfer functions.

In comparison with existing studies, the main contributions of this work are as follows:

(1)\;a method for computer simulation of Bessel, Papoulis, and Legendre filters in continuous time is proposed;

(2)\;it is shown how the proposed method can be extended to modified Bessel filters, Halpern filters, and ultraspherical filters;

(3)\;all necessary two-dimensional nonstationary transfer functions of elementary control system blocks with respect to the trigonometric basis are presented to account for phase delay and to compute the error.

The developed technique for computer simulation of linear filters can be generalized to fractional-order filters. However, for such filters, when using the spectral method, Legendre polynomials should be chosen as the basis, since the explicit formulae for elements of the two-dimensional nonstationary transfer function of a fractional-order integrator/differentiator with respect to this basis are available \cite{Ryb_Comp25, Ryb_MCAP25}.

The remainder of the paper is organized as follows. Section~\ref{secFilters} presents general concepts and definitions related to linear filter theory. The forms of mathematical description of linear filters are also listed there. Bessel, Papoulis, and Legendre filters are considered in Section~\ref{secFilterExamples}, and along with them, modified Bessel filters, Halpern filters, and ultraspherical filters are presented in a more concise form. For each filter family, the transfer function and the two-dimensional nonstationary transfer function are presented. Section~\ref{secSpectral} contains the necessary information about nonstationary spectral characteristics of standard signals and two-dimensional nonstationary transfer functions of elementary blocks with respect to the trigonometric basis for computer simulation of linear filters. The testing methodology and results of the numerical experiment are contained in Section~\ref{secNumerical}. Conclusions are formulated in Section~\ref{secSpConcl}.

\pagebreak

\section{Description of Linear Filters and Problem Statement}\label{secFilters}

In filtering theory, the following types of linear filters are distinguished:
\begin{itemize}
  \item low-pass filters;
  \item high-pass filters;
  \item band-pass filters;
  \item band-stop filters.
\end{itemize}

Low-pass and high-pass filters are characterized by a cutoff frequency $\Omega$, while band-pass and band-stop filters are characterized by lower and upper cutoff frequencies $\Omega_1$ and $\Omega_2$ ($\Omega_1 < \Omega_2$).

An ideal low-pass (high-pass) filter passes harmonic signals with frequencies $\omega \leqslant \Omega$ ($\omega \geqslant \Omega$) and attenuates those with frequencies $\omega > \Omega$ ($\omega < \Omega$). An ideal band-pass (band-stop) filter passes harmonic signals with frequencies $\Omega_1 \leqslant \omega \leqslant \Omega_2$ ($\omega \leqslant \Omega_1$ and $\omega \geqslant \Omega_2$), while those with other frequencies are attenuated.

Traditionally, a linear filter is specified by a transfer function corresponding to a linear time-invariant dynamic system. This is directly related to its purpose and properties. We restrict ourselves to transfer functions of the form
\begin{equation}\label{eqDefH}
  H(s) = \frac{b_m s^m + \ldots + b_1 s + b_0}{a_n s^n + \ldots + a_1 s + a_0}, \ \ \ s \in \mathds{C},
\end{equation}
where $a_0,a_1,\dots,a_n \in \mathds{R}$ and $b_0,b_1,\dots,b_m \in \mathds{R}$ are known coefficients ($a_n \neq 0$ and $b_m \neq 0$), i.e., $H(s)$ is a rational function of a complex variable (ideal filters cannot be described in this way). The number $n$ is called the filter order.

If we denote the input and output signals of the filter by $g(t)$ and $x(t)$, where $t \geqslant 0$ is time, then their Laplace images $G(s)$ and $X(s)$ ($\mathbb{L}$) are related as follows:
\begin{equation}\label{eqInOutTF}
  X(s) = H(s) G(s), \ \ \ G = \mathbb{L} [g], \ \ \ X = \mathbb{L} [x].
\end{equation}

Substituting $s = \mi \omega$, we can pass to the Fourier images $G(\mi \omega)$ and $X(\mi \omega)$ ($\mathbb{F}$):
\[
  X(\mi \omega) = H(\mi \omega) G(\mi \omega), \ \ \ G = \mathbb{F} [g], \ \ \ X = \mathbb{F} [x],
\]
where $\mi$ is the imaginary unit, $\omega \geqslant 0$, and $H(\mi \omega)$ is the frequency response of the filter. Then
\begin{gather*}
  |X(\mi \omega)| = |H(\mi \omega)| |G(\mi \omega)| = A(\omega) |G(\mi \omega)|, \\
  \arg X(\mi \omega) = \arg H(\mi \omega) + \arg G(\mi \omega) = \phi(\omega) + \arg G(\mi \omega).
\end{gather*}

Here, $A(\omega) = |H(\mi \omega)|$ is the amplitude gain of the input harmonic signal with frequency $\omega$ (amplitude frequency response), and $\phi(\omega) = \arg H(\mi \omega)$ determines the phase delay of the output signal relative to the input signal (phase frequency response). The amplitude frequency response and phase frequency response are the most important characteristics of linear filters that perform signal filtering by frequency division \cite{Paa_03}.

The fact that the filter transfer function belongs to the class of rational functions allows the input and output signals to be related by a linear differential equation
\begin{equation}\label{eqODE}
  a_n x^{(n)}(t) + \ldots + a_1 x'(t) + a_0 x(t) = b_m g^{(m)}(t) + \ldots + b_1 g'(t) + b_0 g(t),
\end{equation}
where coefficients on the left-hand side are those of the denominator polynomial of the transfer function, and coefficients on the right-hand side are those of the numerator polynomial.

Along with describing the filter by a linear differential equation for the input--output relationship, we can use an integral transform with a kernel that defines the transfer function and frequency response via the Laplace and Fourier transforms, respectively:
\begin{equation}\label{eqInOutIRF}
  x(t) = \int_0^t k(t-\tau) g(\tau) d\tau, \ \ \ H(s) \colon H = \mathbb{L} [k], \ \ \ H(\mi \omega) \colon H = \mathbb{F} [k],
\end{equation}
where the kernel $k(\eta)$ of the integral operator is called the impulse response (the physical realizability condition $k(\eta) = 0$ for $\eta \leqslant 0$ is used here).

In the present case, the impulse response is represented as a linear combination of functions that can form a fundamental system of solutions of the linear differential equation \eqref{eqODE} under the condition $g(t) = 0$.

Thus, linear filters can be described by any of the three forms presented above. The description via the transfer function and frequency response contains key information about the filter properties, while the description via a linear differential equation or impulse response allows computer simulation of the filter, which usually implies a transition to discrete time \cite{LutTosEva_01, OppSch_14, Gir_17}. However, linear filters can be physically implemented as electrical circuits using resistors, inductors, and capacitors \cite{Lam_82, Dar_TCS84}. The processes occurring in such electrical circuits are continuous-time processes, and they can be modeled without transitioning to discrete time.

Note that the impulse response description is possible even when the filter cannot be specified by the transfer function \eqref{eqDefH} or the linear differential equation \eqref{eqODE}. For example, an ideal low-pass filter with cutoff frequency $\Omega$ has the impulse response $k(\eta) = 2 \Omega \mathop{\mathrm{sinc}} 2 \Omega \eta$, where $\mathop{\mathrm{sinc}} x$ is the (normalized) cardinal sine:
\[
  \mathop{\mathrm{sinc}} x = \left\{ \begin{aligned}
    & \frac{\sin \pi x}{\pi x} & & \text{for} ~ x \neq 0 \\
    & 1 & & \text{for} ~ x = 0,
  \end{aligned} \right.
\]
but such a filter cannot be physically realized. The same remark applies to ideal high-pass, band-pass, and band-stop filters.

Therefore, as filters, linear dynamic systems are used whose amplitude frequency responses approximate the amplitude frequency response of an ideal filter. Since the filter quality can be assessed by various criteria, different filters are used, each with its own advantages and disadvantages.

For computer simulation of linear filters, the spectral method for the analysis of linear control systems can be used \cite{RybShe_FraOp26}. Its application assumes that the input and output signals are represented by ordered sets of expansion coefficients in a chosen basis, and the filter itself is specified by a two-dimensional nonstationary transfer function, which can be readily obtained from the transfer function \cite{SolSemPeshNed_79}.

In the analysis of linear control systems, one usually restricts the time interval on which the input and output signals are considered (this pertains to the analysis of output processes, not to stability analysis). Therefore, below we assume that $t \in \mathds{T} = [0,T]$. The signals $g(t)$ and $x(t)$ are given as expansions in the basis $\BasisT$ of the space $\LPT$, i.e., the space of square-integrable functions \cite{Bal_80}. To avoid complicating the presentation, we suppose that $\BasisT$ is an orthonormal basis. Then
\begin{equation}\label{eqSpInvert}
  g(t) = \sum\limits_{i=0}^\infty G_i q(i,t), \ \ \ x(t) = \sum\limits_{i=0}^\infty X_i q(i,t),
\end{equation}
if $g,x \in \LPT$, where the expansion coefficients $G_i$ and $X_i$ are given by
\[
  G_i = \int_\mathds{T} q(i,t) g(t) dt, \ \ \ X_i = \int_\mathds{T} q(i,t) x(t) dt, \ \ \ i = 0,1,2,\dots
\]

These expansion coefficients contain all information about the functions $g(t)$ and $x(t)$, i.e., the input and output signals are recovered up to equivalence in $\LPT$ from their expansion coefficients. This allows replacing signal transformations by transformations of their expansion coefficients.

For convenience, we define the nonstationary spectral characteristic of a signal as the infinite column matrix of its expansion coefficients in the basis $\BasisT$. Consequently, the input and output signals are associated with their images $G$ and $X$, obtained via the spectral transform ($\mathbb{S}$). These images are infinite column matrices with elements $G_i$ and $X_i$, respectively. Obviously, a linear transformation for an infinite column matrix is given by an infinite matrix. This is the matrix representation of the integral operator with kernel $k(\eta)$ considered as a function of two variables, i.e., $k(t-\tau)$. Such a matrix is called the two-dimensional nonstationary transfer function and denoted in general by $W$. Its elements are defined as follows:
\[
  W_{ij} = \int_{\mathds{T}^2} k(t-\tau) q(i,t) q(j,\tau) dt d\tau, \ \ \ i,j = 0,1,2,\dots
\]

Then
\begin{equation}\label{eqInOutSp}
  X = W G, \ \ \ G = \mathbb{S} [g], \ \ \ X = \mathbb{S} [x],
\end{equation}
which corresponds to formula \eqref{eqInOutTF} in the sense that the application of the Laplace transform and the spectral transform allows passing from differential and integral relations to algebraic ones.

Each linear filter specified by the transfer function \eqref{eqDefH} or the linear differential equation \eqref{eqODE} has its own impulse response \eqref{eqInOutIRF}. However, to find the two-dimensional nonstationary transfer function, it is not necessary to derive the impulse response, which greatly simplifies the use of the spectral method \cite{SolSemPeshNed_79}.

The two-dimensional nonstationary transfer function $W$ is expressed in terms of coefficients $a_0,a_1,\dots,a_n$ and $b_0,b_1,\dots,b_m$ \cite{SolSemPeshNed_79}:
\begin{equation}\label{eqDefW}
  W = (a_n P^n + \ldots + a_1 P + a_0 E)^{-1} (b_m P^m + \ldots + b_1 P + b_0 E),
\end{equation}
where $P$ is the two-dimensional nonstationary transfer function of the differentiator, i.e., the linear system described by the transfer function $H(s) = s$ and the equation $x(t) = g'(t)$, and $E$ is the infinite identity matrix.

In some cases it is convenient to regard $W$ as a function of $P$, i.e., $W = W(P)$, additionally setting $P^0 = E$ (analogous to $s^0 = 1$).

Thus, instead of the complicated transformation
\begin{gather*}
  \text{transfer function} \ \ \ \longrightarrow \ \ \ \text{impulse response} \\
  \longrightarrow \ \ \ \text{two-dimensional nonstationary transfer function}
\end{gather*}
we use the trivial transition
\[
  \text{transfer function} \ \ \ \longrightarrow \ \ \ \text{two-dimensional nonstationary transfer function}
\]
provided that the infinite matrix $P$ is known.

We can also use the two-dimensional nonstationary transfer function $P^{-1}$ of the integrator, i.e., the linear system described by the transfer function $H(s) = 1/s$ and the differential equation $x'(t) = g(t)$. The infinite matrices $P$ and $P^{-1}$ are mutually inverse \cite{SolSemPeshNed_79}:
\[
  P P^{-1} = P^{-1} P = E,
\]
which corresponds to the identity for transfer functions of the differentiator and integrator: $s \cdot 1/s = 1$. In fact, we can regard $W$ as a function of $P^{-1}$, i.e., $W = W(P^{-1})$.

For the numerical experiment in this paper, the trigonometric basis is used. The explicit formulae for basis functions and elements of two-dimensional nonstationary transfer functions $P$ and $P^{-1}$ are given in Section~\ref{secSpectral}.

Further, we denote the numerator and denominator of the transfer function $H(s)$ by $M(s)$ and $D(s)$, respectively. The polynomials $M(s)$ and $D(s)$ are defined on the complex plane $\mathds{C}$, but they can be used for elements of other sets, such as linear operators or their matrix representations, since addition, multiplication by a constant, and integer powers are defined for them.

Indeed, $M(p)$ and $D(p)$ are differential operators corresponding to the right-hand and left-hand sides of the linear differential equation \eqref{eqODE} if $p$ is the time-differentiation operator. Then $M(P)$ and $D(P)$ are their matrix representations:
\[
  W(P) = D^{-1}(P) M(P),
\]
where
\[
  D(P) = a_n P^n + \ldots + a_1 P + a_0 E, \ \ \ M(P) = b_m P^m + \ldots + b_1 P + b_0 E.
\]

To express $M(P)$ and $D(P)$, we can use various representations for polynomials $M(s)$ and $D(s)$. For example, if the roots of these polynomials are given explicitly, then it is convenient to represent $M(s)$ and $D(s)$ in factored form. Then $M(P)$ and $D(P)$ and, consequently, the two-dimensional nonstationary transfer function $W(P)$ can also be conveniently written in factored form. This approach was used for Butterworth, Linkwitz--Riley, and Chebyshev filters in \cite{RybShe_FraOp26}.

If polynomials are defined by a recurrence relation, then $M(P)$ and $D(P)$ can be written analogously. For instance, if $\varphi_n(s)$, $n = 0,1,2,\dots$, are orthogonal polynomials satisfying the recurrence relation
\[
  \varphi_{n+1}(s) = (\alpha_n s + \beta_n) \varphi_n(s) + \gamma_n \varphi_{n-1}(s), \ \ \ n = 1,2,\dots,
\]
where $\{\alpha_n\}$, $\{\beta_n\}$, and $\{\gamma_n\}$ are known numerical sequences, then
\[
  \varphi_{n+1}(P) = (\alpha_n P + \beta_n E) \varphi_n(P) + \gamma_n \varphi_{n-1}(P), \ \ \ n = 1,2,\dots
\]

The transfer functions for different types of linear filters can be expressed through the transfer function of a low-pass filter with cutoff frequency equal to unity. Suppose $H(s)$ is the transfer function of such a filter. Then transfer functions of low-pass and high-pass filters with cutoff frequency $\Omega$ have the form
\[
  H_\Omega(s) = H(s/\Omega) \ \ \ \text{and} \ \ \ H^\Omega(s) = H(\Omega/s),
\]
and transfer functions of band-pass and band-stop filters with cutoff frequencies $\Omega_1$ and $\Omega_2$ are given by
\[
  H_{\Omega_1,\Omega_2}(s) = H(\Omega_1/s) H(s/\Omega_2) \ \ \ \text{and} \ \ \ H^{\Omega_1,\Omega_2}(s) = H(s/\Omega_1) + H(\Omega_2/s).
\]

The relationships between transfer functions and two-dimensional nonstationary transfer functions imply that two-dimensional nonstationary transfer functions of low-pass and high-pass filters with cutoff frequency $\Omega$ are expressed as
\[
  W_\Omega(P) = W(P \, \Omega^{-1}) \ \ \ \text{and} \ \ \ W^\Omega(P) = W(\Omega P^{-1}),
\]
and two-dimensional nonstationary transfer functions of band-pass and band-stop filters with cutoff frequencies $\Omega_1$ and $\Omega_2$ have the form
\[
  W_{\Omega_1,\Omega_2}(P) = W(\Omega_1 P^{-1}) W(P \, \Omega_2^{-1}) \ \ \ \text{and} \ \ \ W^{\Omega_1,\Omega_2}(P) = W(P \, \Omega_1^{-1}) + W(\Omega_2 P^{-1}),
\]
where $W(P)$ is the two-dimensional nonstationary transfer function of the low-pass filter with cutoff frequency equal to unity.

Taking these properties into account, in the general description of filters, one usually restricts attention to low-pass filters with cutoff frequency equal to unity.

The problem is to describe Bessel, Papoulis, and Legendre filters based on the spectral method for the analysis of linear control systems. The main goal of such a description is to develop a new technique for computer simulation of linear filters. Previously, it was applied only to Butterworth, Linkwitz--Riley, and Chebyshev filters.

\begin{remark}\label{remSpectral}~\par
1.\;This section presents four forms of mathematical description of linear filters as linear time-invariant dynamic systems. Along with time-invariant systems, we can also consider time-varying (nonstationary) systems. They cannot be represented by transfer functions, but can be described by linear differential equations with variable coefficients. In the most general form, the input and output signals of a linear dynamic system are related by
\[
  x(t) = \int_0^t k(t,\tau) g(\tau) d\tau,
\]
where the impulse response, i.e., the kernel $k(t,\tau)$ of the integral operator, depends on two variables and satisfies the condition $k(t,\tau) = 0$ for $t \leqslant \tau$.

In this situation, the transition to the spectral domain requires finding the two-dimensional nonstationary transfer function $W$ with elements
\[
  W_{ij} = \int_{\mathds{T}^2} k(t,\tau) q(i,t) q(j,\tau) dt d\tau, \ \ \ i,j = 0,1,2,\dots,
\]
while expression~\eqref{eqInOutSp} remains valid.

2.\;Some clarifications regarding the terminology used in the spectral form of mathematical description are given in \cite{RybShe_FraOp26} and are not repeated here. In particular, the paper \cite{RybShe_FraOp26} includes explanations about nonstationarity and describes a different system of terms used in scientific publications \cite{BagMikPanRyb_Springer20, Ryb_Springer22}.
\end{remark}

For computer simulation of linear filters, two-dimensional nonstationary transfer functions $P$ or $P^{-1}$ are sufficient. However, for a complete analysis of the proposed approach, two additional two-dimensional nonstationary transfer functions are required. The first one, $S^\tau$, corresponds to a pure time shift element by $\tau$ (needed to account for phase delay of the output signal without returning to the time domain). The second one, $A^\theta$, corresponds to an amplifier with gain $a(t) = \chi_{[0,\theta]}(t)$, where $\chi_{[0,\theta]}(t)$ is the indicator of the set $[0,\theta]$ (its purpose is to restrict the interval $[0,T]$ to $[0,\theta]$, $0 < \theta \leqslant T$, for error computation). If $\tau = 0$, then $S^\tau = E$, and if $\theta = T$, then $A^\theta = E$, where $E$ is the infinite identity matrix as before.

Note that all the considered blocks are elementary blocks of control systems. The explicit formulae for elements of two-dimensional nonstationary transfer functions $S^\tau$ and $A^\theta$ defined with respect to the  trigonometric basis are also given in Section~\ref{secSpectral}.

\section{Bessel, Papoulis, and Legendre Filters}\label{secFilterExamples}

In this section, we successively consider three families of low-pass filters (filters in each family differ by order):

1) Bessel filters (BS);

2) Papoulis filters (P);

3) Legendre filters (L),

\noindent
where the short notation given in parentheses is used as a superscript for the transfer function and the two-dimensional nonstationary transfer function.

For each filter family, the following information is provided: the transfer function $H(s)$ for the cutoff frequency $\hat \Omega = 1$ and the corresponding two-dimensional nonstationary transfer function $W(P)$. Additionally, families of filters related to the above are described. They are synthesized based on modified Bessel polynomials and Jacobi polynomials (or Gegenbauer polynomials):

1) modified Bessel filters (BM);

2) Halpern filters (H);

3) ultraspherical filters (U).

The formulae that allow one to express transfer functions and two-dimensional nonstationary transfer functions of a high-pass filter, as well as band-pass and band-stop filters, are given in Section~\ref{secFilters}.

The dependence of the transfer function and the two-dimensional nonstationary transfer function on the filter order and numerical parameters is not indicated to avoid complicating the notation. Here, we assume that the filter order $n$ can be uniquely identified from the given formulae.

By analogy with \cite{RybShe_FraOp26}, to visualize the characteristic polynomials of some linear filters, i.e., denominators of transfer functions, we use the ``domain coloring'' technique based on the HLS color model \cite{Weg_12}:
\[
  \text{``Hue''} \propto \arg H(s), \ \ \ \text{``Lightness''} \propto \arctan |H(s)|, \ \ \ \text{``Saturation''} = 100\%.
\]

\subsection{Bessel Filters}

The transfer function of the Bessel filter of order $n$ is given by
\begin{equation}\label{eqTFBS}
  H^\bs(s) = \frac{\theta_n(0)}{\theta_n(s)},
\end{equation}
where $\theta_n(s)$ is a polynomial of degree $n$, called the reverse Bessel polynomial. The numerator of the transfer function contains a normalizing factor, which ensures the standard condition, namely the transfer function equals unity when $s = 0$.

The explicit formula defining Bessel polynomials \cite{KraFri_TAMS49} is
\[
  y_n(s) = \sum\limits_{k = 0}^n a_{nk} s^k,
\]
where the coefficients $a_{nk}$ are given by
\[
  a_{nk} = \frac{(n+k)!}{2^k \, (n-k)! \, k!}, \ \ \ k = 0,1,\dots,n,
\]
and $n$ is the degree of the polynomial.

In formula \eqref{eqTFBS}, polynomials with reversed coefficients (reverse Bessel polynomials) are used:
\[
  \theta_n(s) = \sum\limits_{k = 0}^n a_{nk} s^{n-k} = \sum\limits_{k = 0}^n a_{n,n-k} s^k,
\]
related to $y_n(s)$ by
\[
  \theta_n(s) = s^n y_n \biggl( \frac{1}{s} \biggr).
\]

In the context of applying the spectral method, Butterworth, Linkwitz--Riley, and Chebyshev filters were previously considered \cite{RybShe_FraOp26}. Their transfer functions are defined on the basis of Butterworth and Chebyshev polynomials, which are most conveniently specified by their roots (the corresponding transfer functions are completely determined by zeros and poles, taking their multiplicities into account). However, for Bessel polynomials there are no explicit formulae for the roots except for the case $n \leqslant 4$, when known rules for finding the roots of polynomials of degrees 2, 3, and 4 can be applied. Nevertheless, it has been proved that all roots of polynomials $y_n(s)$ and $\theta_n(s)$ have negative real parts \cite{Gro_78}, i.e., all poles of the transfer function $H^\bs(s)$ lie in the left half-plane $\{s \colon \Re s < 0\}$, and it defines an asymptotically stable linear dynamic system.

Thus, using formula \eqref{eqDefW} and taking into account the equality
\[
  \theta_n(0) = a_{nn} = \frac{(2n)!}{2^n n!} = (2n-1)!!,
\]
we obtain the two-dimensional nonstationary transfer function of the Bessel filter:
\begin{equation}\label{eqNTFBW1}
  W^\bs(P) = \frac{(2n)!}{2^n n!} \biggl( \, \sum\limits_{k = 0}^n a_{nk} P^{n-k} \biggr)^{-1} = (2n-1)!! \biggl( \, \sum\limits_{k = 0}^n a_{n,n-k} P^k \biggr)^{-1}.
\end{equation}

The polynomials $y_n(s)$ and $\theta_n(s)$ satisfy the recurrence relations
\begin{gather*}
  y_{n+1}(s) = (2n+1) s y_n(s) + y_{n-1}(s), \ \ \ \theta_{n+1}(s) = (2n+1) \theta_n(s) + s^2 \theta_{n-1}(s), \\
  y_0(s) = \theta_0(s) = 1, \ \ \ y_1(s) = \theta_1(s) = s+1, \ \ \ n = 1,2,\dots
\end{gather*}

Based on them, we can propose equivalent relations for the two-dimensional nonstationary transfer function of the Bessel filter:
\begin{equation}\label{eqNTFBW2}
  W^\bs(P) = (2n-1)!! \, \theta_n^{-1}(P) \ \ \ \text{or} \ \ \ W^\bs(P) = (2n-1)!! \, y_n^{-1}(P^{-1}) P^{-n},
\end{equation}
where
\[
  y_{n+1}(P^{-1}) = (2n+1) P^{-1} y_n(P^{-1}) + y_{n-1}(P^{-1}), \ \ \ y_0(P^{-1}) = E, \ \ \ y_1(P^{-1}) = P^{-1} + E
\]
and
\[
  \theta_{n+1}(P) = (2n+1) \theta_n(P) + P^2 \theta_{n-1}(P), \ \ \ \theta_0(P) = E, \ \ \ \theta_1(P) = P + E.
\]

\textbf{Example 1.} {\itshape Consider the Bessel filter of order $n = 3$. The Bessel polynomials up to degree $n = 3$ are obtained by the recurrence relation:
\begin{gather*}
  y_0(s) = 1, \ \ \ y_1(s) = s+1, \ \ \ y_2(s) = 3s y_1(s) + y_0(s) = 3s^2 + 3s + 1, \\
  y_3(s) = 5 s y_2(s) + y_1(s) = 15s^3 + 15s^2 + 6s + 1.
\end{gather*}

Reversing the order of coefficients gives reverse Bessel polynomials:
\[
  \theta_0(s) = 1, \ \ \ \theta_1(s) = s+1, \ \ \ \theta_2(s) = s^2 + 3s + 3, \ \ \ \theta_3(s) = s^3 + 6s^2 + 15s + 15.
\]

Consequently, the transfer function of the Bessel filter of order $n = 3$ is
\[
  H^\bs(s) = \frac{5!!}{\theta_3(s)} = \frac{15}{s^3 + 6s^2 + 15s + 15},
\]
and the corresponding two-dimensional nonstationary transfer function is expressed as
\[
  W^\bs(P) = 5!! \, \theta_3^{-1}(P) = 15 \, (P^3 + 6P^2 + 15P + 15E)^{-1},
\]
since
\[
  \theta_3(P) = P^3 + 6P^2 + 15P + 15E,
\]
where $P$ is the two-dimensional nonstationary transfer function of the differentiator.

The same result is obtained using the recurrence relation for reverse Bessel polynomials:
\begin{gather*}
  \theta_0(P) = E, \ \ \ \theta_1(P) = P + E, \ \ \ \theta_2(P) = 3 \theta_1(P) + P^2 \theta_0(P) = P^2 + 3P + 3E, \\
  \theta_3(P) = 5 \theta_2(P) + P^2 \theta_1(P) = P^3 + 6P^2 + 15P + 15E.
\end{gather*}

Figure~\ref{picBesselPoly} shows the characteristic polynomial of the Bessel filter (the denominator of the transfer function) on a fragment of the complex plane $\{s \colon \Re s,\Im s \in [-3,3]\}$, filter order $n = 3$. The roots of the characteristic polynomial (poles of the transfer function) appear as three black dots.

\begin{figure}[ht]
  \centering
  \ifnum \showfigures = 1
  \includegraphics[scale = 1]{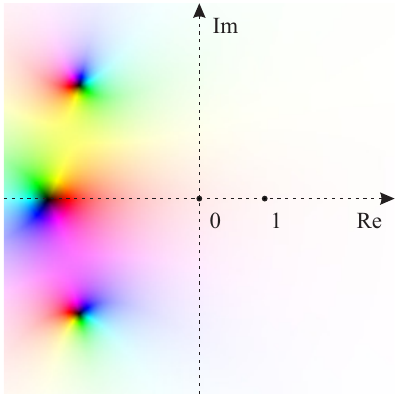}
  \fi
  \caption{Characteristic polynomial of the Bessel filter ($n = 3$)}\label{picBesselPoly}
\end{figure}

An equivalent representation of the two-dimensional nonstationary transfer function of the Bessel filter of order $n = 3$ uses the two-dimensional nonstationary transfer function of the integrator $P^{-1}$:
\[
  W^\bs(P) = 5!! \, y_3^{-1}(P^{-1}) P^{-3} = 15 \, (15P^{-3} + 15P^{-2} + 6P^{-1} + E)^{-1} P^{-3}.
\]

}

We can consider a wider class of filters that includes Bessel filters as a special case, namely generalized Bessel filters \cite{JohnJohnBouSto_TCS76}.

To describe them, we introduce one-parameter generalized Bessel polynomials and their reverse counterparts:
\[
  y_n^\alpha(s) = \sum\limits_{k = 0}^n a_{nk}^\alpha s^k, \ \ \ \theta_n^\alpha(s) = \sum\limits_{k = 0}^n a_{nk}^\alpha s^{n-k} = \sum\limits_{k = 0}^n a_{n,n-k}^\alpha s^k,
\]
for which the coefficients $a_{nk}^\alpha$ are given by
\[
  a_{nk}^\alpha = \frac{n!}{2^k \, k! \, (n-k)!} \, (n + \alpha + 1)^{\overline{k}}, \ \ \ k = 0,1,\dots,n,
\]
where $(n + \alpha + 1)^{\overline{k}}$ is the rising factorial:
\[
  x^{\overline{k}} = x (x+1) \ldots (x+k-1),
\]
and $\alpha > -2n$ is a parameter.

The transfer function of the generalized Bessel filter of order $n$ is defined as
\[
  H^\bm(s) = \frac{\theta_n^\alpha(0)}{\theta_n^\alpha(s)};
\]
consequently, we obtain the two-dimensional nonstationary transfer function of the generalized Bessel filter:
\[
  W^\bm(P) = \frac{(n + \alpha + 1)^{\overline{n}}}{2^n}  \biggl( \, \sum\limits_{k = 0}^n a_{nk}^\alpha P^{n-k} \biggr)^{-1} = \frac{(n + \alpha + 1)^{\overline{n}}}{2^n}  \biggl( \, \sum\limits_{k = 0}^n a_{n,n-k}^\alpha P^k \biggr)^{-1},
\]
since
\[
  \theta_n^\alpha(0) = a_{nn}^\alpha = \frac{(n + \alpha + 1)^{\overline{n}}}{2^n}.
\]

The case $\alpha = 0$ corresponds to Bessel polynomials and the Bessel filter considered above.

For generalized Bessel polynomials, we can also use recurrence relations, carrying them over to the spectral domain \cite{Gro_78}. They can be expressed in terms of Bessel polynomials. There also exist two- and three-parameter generalized Bessel polynomials on the basis of which linear filters have been proposed \cite{JohnJohnBouSto_TCS76}. The derivation of two-dimensional nonstationary transfer functions for them is analogous to the cases considered and is not given here.

\subsection{Papoulis Filters}

The construction of the transfer function of the Papoulis filter is more laborious than that of the Bessel filter.

First, we recall the explicit formulae for standard Legendre polynomials $p_n(x)$ \cite{KornKorn_00}:
\begin{equation}\label{eqLegExp}
  p_n(x) = \frac{1}{2^n} \sum\limits_{k = 0}^{\lfloor n/2 \rfloor} \frac{(-1)^k(2n-2k)!}{k! \, (n-k)! \, (n-2k)!} \, x^{n-2k}, \ \ \ n = 0,1,2,\dots,
\end{equation}
where $\lfloor \,\cdot\, \rfloor$ is the floor function. However, in many situations, it is simpler to use the recurrence relation defining Legendre polynomials:
\[
  (n+1) p_{n+1}(x) = (2n+1) x p_n(x) - n p_{n-1}(x), \ \ \ p_0(x) = 1, \ \ \ p_1(x) = x, \ \ \ n = 1,2,\dots
\]

Next, define the function $L_{2n}(z)$ as follows ($L_{2n}(z)$ is a polynomial of degree $2n$):
\[
  L_{2n}(z) = \int_{-1}^{2z^2-1} (x+1)^{(n+1)\,(\mathrm{mod}~2)} \biggl( \, \sum\limits_{k=0}^m \alpha_k p_k(x) \biggr)^2 dx,
\]
where $m = \lfloor (n-1)/2 \rfloor$. Coefficients $\alpha_k$ are determined depending on whether the filter order $n$ is even or odd. If $n$ is odd, then
\[
  \alpha_k = \frac{1}{\sqrt{2}} \, \frac{2k+1}{m+1}, \ \ \ k = 0,1,\dots,m,
\]
and if $n$ is even, then
\[
  \alpha_k = \frac{2k+1}{\sqrt{(m+1)(m+2)}},
\]
where the latter formula is used for values $k = 0,1,\dots,m$ having the same parity as $m$, while the other coefficients are zero.

For even-order filters, there is another formula for $L_{2n}(z)$ using the derivative of the Legendre polynomial of degree $m+1$ \cite{Pap_PIRE59}, but it is equivalent to the above formula.

The next step is to find the roots of the algebraic equation $L_{2n}(\mi s) = -1$ of degree $2n$, from which those with negative real parts are selected. From them a new polynomial $l_n(s)$ of degree $n$ is constructed with the additional condition $l_n(0) = 1$, i.e., spectral factorization of the polynomial $1 + L_{2n}(z)$ is used:
\[
  1 + L_{2n}(z) = l_n(\mi z) l_n(-\mi z).
\]

Finding the roots of the equation $L_{2n}(\mi s) = -1$ in general reduces to various approximate methods. A survey of such methods can be found in \cite{ManMar_26}.

The transfer function of the Papoulis filter of order $n$ is expressed via the polynomial $l_n(s)$ (a Hurwitz polynomial):
\begin{equation}\label{eqTFP}
  H^\pp(s) = \frac{1}{l_n(s)}.
\end{equation}

Using the representation of $l_n(s)$ as
\[
  l_n(s) = \sum\limits_{k = 0}^n \lambda_{nk} s^k,
\]
where the coefficients $\lambda_{nk}$ are determined by the roots of the equation $L_{2n}(\mi s) = -1$ with negative real parts, as well as by condition $l_n(0) = \lambda_{n0} = 1$, we obtain the two-dimensional nonstationary transfer function of the Papoulis filter:
\begin{equation}\label{eqNTFP}
  W^\pp(P) = l_n^{-1}(P) = \biggl( \, \sum\limits_{k = 0}^n \lambda_{nk} P^k \biggr)^{-1}.
\end{equation}

\textbf{Example 2.} {\itshape Consider the Papoulis filter of even order $n = 4$. In this case, $m = \lfloor 3/2 \rfloor = 1$ is odd, so $\alpha_0 = 0$, $\alpha_1 = 3/\sqrt{6}$, and
\[
  \sum\limits_{k=0}^1 \alpha_k p_k(x) = \alpha_1 p_1(x) = \frac{3}{\sqrt{6}} \, x.
\]

Consequently,
\[
  L_8(z) = \frac{3}{2} \int_{-1}^{2z^2-1} (x+1) x^2 dx = 6 z^8 - 8 z^6 + 3 z^4
\]
and
\[
  1 + L_8(\mi s) = 6s^8 + 8s^6 + 3s^4 + 1.
\]

The equation $L_8(\mi s) = -1$ has eight roots (four complex conjugate pairs): $\pm 0.549743 \pm 0.358572 \, \mi$ and $\pm 0.231689 \pm 0.945511 \, \mi$. To find the roots, the Laguerre method was used; however, they can be determined exactly by applying Ferrari's method \cite{KornKorn_00} to the quartic equation $6w^4 + 8w^3 + 3w^2 + 1 = 0$, $w = s^2$.

From the four roots $-0.549743 \pm 0.358572 \, \mi$ and $-0.231689 \pm 0.945511 \, \mi$ we obtain a polynomial of degree $n = 4$:
\begin{align*}
  & (s + 0.549743 + 0.358572 \, \mi)(s + 0.549743 - 0.358572 \, \mi) \\
  & \ \ \ \ \ \ {} \times (s + 0.231689 + 0.945511 \, \mi)(s + 0.231689 - 0.945511 \, \mi) \\
  & \ \ \ = (s^2 + 1.099486 s + 0.430791)(s^2 + 0.463378 s + 0.947671) \\
  & \ \ \ = s^4 + 1.562864 s^3 + 1.887939 s^2 + 1.241570 s + 0.408248,
\end{align*}
which should be divided by the constant term 0.408248. Then
\[
  l_4(s) = 2.449490 s^4 + 3.828220 s^3 + 4.624487 s^2 + 3.041213 s + 1 \ \ \ \text{and} \ \ \ l_4(0) = 1.
\]

Thus, we find the transfer function of the Papoulis filter of order $n = 4$:
\[
  H^\pp(s) = \frac{1}{l_4(s)} = \frac{1}{2.449490 s^4 + 3.828220 s^3 + 4.624487 s^2 + 3.041213 s + 1},
\]
and the corresponding two-dimensional nonstationary transfer function is
\[
  W^\pp(P) = l_4^{-1}(P) = (2.449490 P^4 + 3.828220 P^3 + 4.624487 P^2 + 3.041213 P + E)^{-1},
\]
where $P$ is the two-dimensional nonstationary transfer function of the differentiator.

The characteristic polynomial of the Papoulis filter (denominator of the transfer function) is shown in Figure~\ref{picPapoulisPoly} (fragment of the complex plane $\{s \colon \Re s,\Im s \in [-2,2]\}$, filter order $n = 4$). Four black dots correspond to the roots of the characteristic polynomial (poles of the transfer function).

\begin{figure}[ht]
  \centering
  \ifnum \showfigures = 1
  \includegraphics[scale = 1]{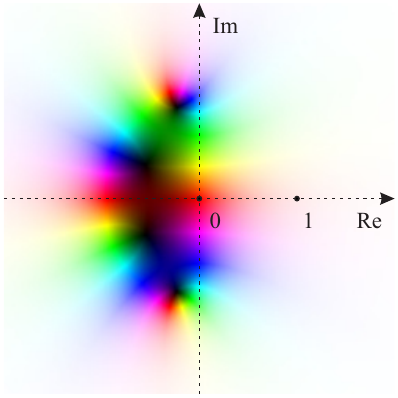}
  \fi
  \caption{Characteristic polynomial of the Papoulis filter ($n = 4$)}\label{picPapoulisPoly}
\end{figure}

If we use the approach for representing the two-dimensional nonstationary transfer function of a linear filter from \cite{RybShe_FraOp26}, we can write the equivalent formula
\[
  W^\pp(P) = 0.408248 \bigl( (P^2 + 1.099486 P + 0.430791 E)(P^2 + 0.463378 P + 0.947671 E) \bigr)^{-1}.
\]

}

\begin{remark}\label{remPapoulis}~\par
1.\;A more general case involves the parameter $\varepsilon$ controlling the filter attenuation coefficient. The difference from the considered version is that the roots of the algebraic equation $\varepsilon^2 L_{2n}(\mi s) = -1$ are found. This changes coefficients of the polynomial $l_n(s)$:
\[
  1 + \varepsilon^2 L_{2n}(z) = l_n(\mi z) l_n(-\mi z),
\]
but the condition $l_n(0) = 1$ must still hold. From coefficients of $l_n(s)$, it is straightforward to obtain the two-dimensional nonstationary transfer function of the Papoulis filter taking into account the parameter $\varepsilon$.

2.\;For low orders $n$, the roots of the algebraic equation $L_{2n}(\mi s) = -1$ and transfer functions $H^\pp(s)$ are known \cite{Pap_PIRE58, Pap_PIRE59}.
\end{remark}

Further, we consider Halpern filters \cite{Hal_TCT69}, which are synthesized similarly to Papoulis filters. Define the polynomial $T_{2n}(z)$ of degree $2n$ by
\[
  T_{2n}(z) = \int_0^z x u_{n-1}^2(x) dx,
\]
where
\[
  u_{2m+l}(x) = \sqrt{4m+2(1+l)} \sum\limits_{k = 0}^m \frac{(-1)^{m-k} (m+k+l)!}{(m-k)! \, k! \, (k+l)!} \, x^{2k+1},
\]
with $m = \lfloor (n-1)/2 \rfloor$ and $l = (n-1)\,(\mathrm{mod}~2)$. It is related to Legendre polynomials; in particular, nonzero coefficients of $u_{2m+l}(x)$ for $l = 1$ differ from coefficients of the Legendre polynomial of degree $m$ defined on $[0,1]$ by a constant factor (Legendre polynomials $p_n(x)$ given by \eqref{eqLegExp} are defined on $[-1,1]$).

To synthesize the Halpern filter, we should find the roots of the algebraic equation $T_{2n}(\mi s) = -1$ of degree $2n$ and select those with negative real parts. Then a new polynomial $\vartheta_n(s)$ of degree $n$ with the additional condition $\vartheta_n(0) = 1$ is constructed (spectral factorization of $1 + T_{2n}(z)$):
\[
  1 + T_{2n}(z) = \vartheta_n(\mi z) \vartheta_n(-\mi z).
\]

It remains to obtain the transfer function of the Halpern filter of order $n$:
\[
  H^\ha(s) = \frac{1}{\vartheta_n(s)} = \biggl( \sum\limits_{k = 0}^n \theta_{nk} s^k \biggr)^{-1},
\]
where $\theta_{nk}$ are the coefficients of $\vartheta_n(s)$. Then the corresponding two-dimensional nonstationary transfer function is
\[
  W^\ha(P) = \vartheta_n^{-1}(P) = \biggl( \, \sum\limits_{k = 0}^n \theta_{nk} P^k \biggr)^{-1}.
\]

For low orders $n$, polynomials $u_{n-1}(x)$ and $T_{2n}(z)$ are given in \cite{Hal_TCT69}. In a more general setting, we can use the parameter $\varepsilon$ controlling the attenuation coefficient (see Remark~\ref{remPapoulis}). Also, linear filters generalizing Halpern filters are used \cite{AttDang_ProcIEEE78}. For them, we can obtain two-dimensional nonstationary transfer functions using a similar approach.

\subsection{Legendre Filters}

The synthesis of filters called Legendre filters involves the use of modified associated Legendre polynomials. They are defined as
\[
  p_n^{(m)}(x) = C_{n,m} \, \frac{d^m p_{n+m}(x)}{dx^m}, \ \ \ m = 0,1,2,\dots,
\]
where $p_{n+m}(x)$ is the Legendre polynomial of degree $n+m$ given by \eqref{eqLegExp}, and $C_{n,m}$ is a normalizing factor chosen from the condition $\max\limits_{x \in [-1,1]} |p_n^{(m)}(x)| = 1$:
\[
  C_{n,m} = p_{n,m}(1), \ \ \ p_{n,m}(x) = \frac{d^m p_{n+m}(x)}{dx^m}.
\]

The polynomials $p_{n,m}(x)$ are called associated Legendre polynomials, while $p_n^{(m)}(x)$ are modified associated Legendre polynomials. Note that in mathematical handbooks, associated Legendre polynomials (or more correctly, associated Legendre functions) are defined differently \cite{KornKorn_00, BatErd_53, PolMan_07}. However, in this paper we use the terminology adopted in the literature on linear filter theory \cite{KuDru_JFI62, Paa_03}.

Based on modified associated Legendre polynomials, the transfer function is synthesized. Next, from the roots of the algebraic equation $(p_n^{(m)}(\mi s))^2 = -1$ of degree $2n$, those with negative real parts are selected. Then a new polynomial $\varkappa_n(s)$ of degree $n$ is constructed (spectral factorization of $1 + (p_n^{(m)}(z))^2$):
\[
  1 + \bigl( p_n^{(m)}(z) \bigr)^2 = \varkappa_n(\mi z) \varkappa_n(-\mi z), \ \ \ \varkappa_n(0) = 1.
\]

As for Papoulis filters, finding the roots of the equation $(p_n^{(m)}(\mi s))^2 = -1$ generally requires various approximate methods.

The polynomial $\varkappa_n(s)$ (a Hurwitz polynomial) completely determines the transfer function of the Legendre filter of order $n$ for a chosen parameter $m$:
\begin{equation}\label{eqTFL}
  H^\ld(s) = \frac{1}{\varkappa_n(s)}.
\end{equation}

Using the representation
\[
  \varkappa_n(s) = \sum\limits_{k = 0}^n \mu_{nk} s^k,
\]
where the coefficients $\mu_{nk}$ are determined by the roots of the equation $(p_n^{(m)}(\mi s))^2 = -1$ with negative real parts and the condition $\varkappa_n(0) = \mu_{n0} = 1$, we obtain the two-dimensional nonstationary transfer function of the Legendre filter:
\begin{equation}\label{eqNTFL}
  W^\ld(P) = \varkappa_n^{-1}(P) = \biggl( \, \sum\limits_{k = 0}^n \mu_{nk} P^k \biggr)^{-1}.
\end{equation}

\textbf{Example 3.} {\itshape Consider Legendre filters of order $n = 5$ with $m = 1$ and $m = 2$. To find their transfer functions, we need Legendre polynomials of degree 6 and 7:
\[
  p_6(x) = \frac{231 x^6 - 315 x^4 + 105 x^2 - 5}{16}, \ \ \ p_7(x) = \frac{429 x^7 - 693 x^5 + 315 x^3 - 35 x}{16},
\]
and their first and second derivatives (associated Legendre polynomials), respectively:
\begin{align*}
  p_{5,1}(x) & = \frac{dp_6(x)}{dx} = \frac{21 x \, (33 x^4 - 30 x^2 + 5)}{8}, & p_{5,1}(1) & = 21; \\
  p_{5,2}(x) & = \frac{d^2 p_7(x)}{dx^2} = \frac{63 x \, (143 x^4 - 110 x^2 + 15)}{8}, & p_{5,2}(1) & = 378.
\end{align*}

Hence, modified associated Legendre polynomials are
\[
  p_{5}^{(1)}(x) = \frac{p_{5,1}(x)}{p_{5,1}(1)} = \frac{33 x^5 - 30 x^3 + 5 x}{8}, \ \ \
  p_{5}^{(2)}(x) = \frac{p_{5,2}(x)}{p_{5,2}(1)} = \frac{143 x^5 - 110 x^3 + 15 x}{48},
\]
then
\begin{align*}
  1 + \bigl( p_5^{(1)}(\mi s) \bigr)^2 & = \frac{1089}{64} \, s^{10} - \frac{495}{16} \, s^8 + \frac{615}{32} \, s^6 - \frac{75}{16} \, s^4 + \frac{25}{64} \, s^2 + 1, \\
  1 + \bigl( p_5^{(2)}(\mi s) \bigr)^2 & = \frac{20449}{2304} \, s^{10} - \frac{7865}{576} \, s^8 + \frac{8195}{1152} \, s^6 - \frac{275}{192} \, s^4 + \frac{25}{256} s^2 + 1.
\end{align*}

We start with the case $m = 1$. The equation $(p_5^{(1)}(\mi s))^2 = -1$ has ten roots (two real roots and four complex conjugate pairs): $\pm 0.518232$, $\pm 0.412892 \pm 0.578058 \, \mi$, and $\pm 0.153776 \pm 0.950447 \, \mi$ (the Laguerre method was used). From the five roots $-0.518232$, $-0.412892 \pm 0.578058 \, \mi$, and $-0.153776 \pm 0.950447 \, \mi$, we obtain a polynomial of degree $n = 5$:
\begin{align*}
  & (s + 0.518232)(s + 0.412892 + 0.578058 \, \mi)(s + 0.412892 - 0.578058 \, \mi) \\
  & \ \ \ \ \ \ {} \times (s + 0.153776 + 0.950447 \, \mi)(s + 0.153776 - 0.950447 \, \mi) \\
  & \ \ \ = (s + 0.518232)(s^2 + 0.825784 s + 0.710374)(s^2 + 0.307553 s + 0.962806) \\
  & \ \ \ = s^5 + 1.651569 s^4 + 2.272931 s^3 + 1.794231 s^2 + 0.944927 s + 0.242424,
\end{align*}
which should be divided by the constant term 0.242424. Then
\[
  \varkappa_5(s) = 4.125000 s^5 + 6.812722 s^4 + 9.375840 s^3 + 7.401202 s^2 + 3.897824 s + 1 \ \ \ \text{and} \ \ \ \varkappa_5(0) = 1.
\]

Therefore, the transfer function of the Legendre filter of order $n = 5$ with $m = 1$ is
\[
  H^\ld(s) = \frac{1}{\varkappa_5(s)} = \frac{1}{4.125000 s^5 + 6.812722 s^4 + 9.375840 s^3 + 7.401202 s^2 + 3.897824 s + 1},
\]
and the corresponding two-dimensional nonstationary transfer function is
\[
  W^\ld(P) = \varkappa_5^{-1}(P) = (4.125000 P^5 + 6.812722 P^4 + 9.375840 P^3 + 7.401202 P^2 + 3.897824 P + E)^{-1},
\]
or in the equivalent form
\[
  W^\ld(P) = 0.242424 \bigl( (P + 0.518232 E)(P^2 + 0.825784 P + 0.710374 E)(P^2 + 0.307553 P + 0.962806 E) \bigr)^{-1},
\]
where $P$ is the two-dimensional nonstationary transfer function of the differentiator.

A visualization of the characteristic polynomial of the Legendre filter (denominator of the transfer function) is shown in Figure~\ref{picLegendre1Poly} (fragment of the complex plane $\{s \colon \Re s,\Im s \in [-2,2]\}$, filter order $n = 5$, $m = 1$). Five black dots correspond to the roots of the characteristic polynomial (poles of the transfer function).

\begin{figure}[ht]
  \centering
  \ifnum \showfigures = 1
  \includegraphics[scale = 1]{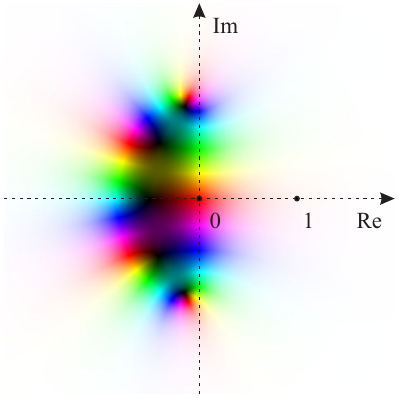}
  \fi
  \caption{Characteristic polynomial of the Legendre filter ($n = 5$, $m = 1$)}\label{picLegendre1Poly}
\end{figure}

Next, we perform similar calculations for the case $m = 2$. The equation $(p_5^{(2)}(\mi s))^2 = -1$ also has ten roots (two real roots and four complex conjugate pairs): $\pm 0.617494$, $\pm 0.494148 \pm 0.579825 \, \mi$, and $\pm 0.185401 \pm 0.949863 \, \mi$ (all roots were found by the Laguerre method). From the five roots $-0.617494$, $-0.494148 \pm 0.579825 \, \mi$, and $-0.185401 \pm 0.949863 \, \mi$, we derive a polynomial of degree $n = 5$:
\begin{align*}
  & (s + 0.617494)(s + 0.494148 + 0.579825 \, \mi)(s + 0.494148 - 0.579825 \, \mi) \\
  & \ \ \ \ \ \ {} \times (s + 0.185401 + 0.949863 \, \mi)(s + 0.185401 - 0.949863 \, \mi) \\
  & \ \ \ = (s + 0.617494)(s^2 + 0.988297 s + 0.761826)(s^2 + 0.370802 s + 0.967788) \\
  & \ \ \ = s^5 + 1.976593 s^4 + 2.722691 s^3 + 2.303880 s^2 + 1.248064 s + 0.335664,
\end{align*}
which should be divided by the constant term 0.335664. Then
\[
  \varkappa_5(s) = 2.979167 s^5 + 5.888602 s^4 + 8.111352 s^3 + 6.863645 s^2 + 3.718191 s + 1 \ \ \ \text{and} \ \ \ \varkappa_5(0) = 1.
\]

Consequently, we find the transfer function of the Legendre filter of order $n = 5$ with $m = 2$:
\[
  H^\ld(s) = \frac{1}{\varkappa_5(s)} = \frac{1}{2.979167 s^5 + 5.888602 s^4 + 8.111352 s^3 + 6.863645 s^2 + 3.718191 s + 1},
\]
and the two-dimensional nonstationary transfer function:
\[
  W^\ld(P) = \varkappa_5^{-1}(P) = (2.979167 P^5 + 5.888602 P^4 + 8.111352 P^3 + 6.863645 P^2 + 3.718191 P + E)^{-1},
\]
or in the equivalent form
\[
  W^\ld(P) = 0.335664 \bigl( (P + 0.617494 E)(P^2 + 0.988297 P + 0.761826 E)(P^2 + 0.370802 P + 0.967788 E) \bigr)^{-1}.
\]

A visualization of the characteristic polynomial of the Legendre filter is shown in Figure~\ref{picLegendre2Poly} (fragment of the complex plane $\{s \colon \Re s,\Im s \in [-2,2]\}$, filter order $n = 5$, $m = 2$).

\begin{figure}[ht]
  \centering
  \ifnum \showfigures = 1
  \includegraphics[scale = 1]{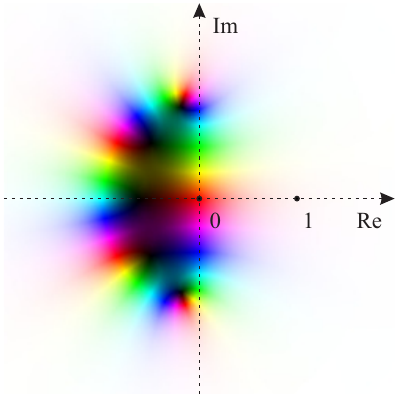}
  \fi
  \caption{Characteristic polynomial of the Legendre filter ($n = 5$, $m = 2$)}\label{picLegendre2Poly}
\end{figure}

}

\begin{remark}\label{remLegendre}~\par
1.\;For Legendre filters, the general case involves the use of the parameter $\varepsilon$ (see Remark~\ref{remPapoulis}) controlling the attenuation coefficient. Then, for the filter synthesis, we should find the roots of the algebraic equation $\varepsilon^2 (p_n^{(m)}(\mi s))^2 = -1$ (similar to Papoulis filters):
\[
  1 + \varepsilon^2 \bigl( p_n^{(m)}(z) \bigr)^2 = \varkappa_n(\mi z) \varkappa_n(-\mi z), \ \ \ \varkappa_n(0) = 1.
\]

The two-dimensional nonstationary transfer function of the Legendre filter taking into account the parameter $\varepsilon$ is specified by coefficients of the polynomial $\varkappa_n(s)$.

2.\;Legendre polynomials are also used for the synthesis of filters described in \cite{ChrSah_EE99}. There, it is assumed that the characteristic polynomial is formed from a product of Legendre polynomials with total degree $n$.
\end{remark}

Legendre filters are generalized as follows: spectral factorization is applied to the polynomial $1 + (f_n^{(\alpha)}(z))^2$ instead of $1 + (p_n^{(m)}(z))^2$, i.e.,
\[
  1 + \bigl( f_n^{(\alpha)}(z) \bigr)^2 = \psi_n(\mi z) \psi_n(-\mi z), \ \ \ \psi_n(0) = 1.
\]

The polynomial $f_n^{(\alpha)}(x)$ is given by (the rising factorial was introduced in the previous subsection)
\[
  f_n^{(\alpha)}(x) = \frac{n!}{\alpha^{\overline{k}}} \, p_n^{(\alpha,\alpha)}(x),
\]
where $p_n^{(\alpha,\alpha)}(x)$ is the Jacobi polynomial (ultraspherical polynomial) \cite{Paa_03, JohnJohn_TCT66} with $\alpha > -1$.

In general, Jacobi polynomials form a two-parameter family of the orthogonal polynomials $p_n^{(\alpha,\beta)}(x)$ \cite{KornKorn_00, BatErd_53, PolMan_07}. Here, we use Jacobi polynomials with $\alpha = \beta$. As a result, the polynomial $f_n^{(\alpha)}(x)$ has the explicit form
\[
  f_n^{(\alpha)}(x) = \sum\limits_{k = 0}^n \frac{(-N)^{\overline{k}} (N+2\alpha+1)^{\overline{k}}}{k! \, (\alpha+1)^{\overline{k}}} \biggl( \frac{1-x}{2} \biggr)^k,
\]
or we can use the recurrence relation
\begin{gather*}
  (n+2\alpha+1) f_{n+1}^{(\alpha)}(x) = (2n+2\alpha+1) x f_n^{(\alpha)}(x) - n f_{n-1}^{(\alpha)}(x), \\
  f_0^{(\alpha)}(x) = 1, \ \ \ f_1^{(\alpha)}(x) = x, \ \ \ n = 1,2,\dots
\end{gather*}

Then the transfer function of the ultraspherical filter of order $n$ is
\[
  H^\us(s) = \frac{1}{\psi_n(s)} = \biggl( \sum\limits_{k = 0}^n \nu_{nk} s^k \biggr)^{-1},
\]
where the coefficients $\nu_{nk}$ are determined by the roots of the algebraic equation $(f_n^{(\alpha)}(\mi s))^2 = -1$ of degree $2n$ with negative real parts and the condition $\psi_n(0) = \nu_{n0} = 1$. Its two-dimensional nonstationary transfer function is
\[
  W^\us(P) = \psi_n^{-1}(P) = \biggl( \, \sum\limits_{k = 0}^n \nu_{nk} P^k \biggr)^{-1}.
\]

Up to a constant factor, the polynomial $f_n^{(\alpha)}(x)$ coincides with the Gegenbauer polynomial of the same degree but with the parameter $\alpha+1/2$. For $\alpha = 0$, such polynomials reduce to Legendre polynomials; for $\alpha = m = 1,2,\dots$, to associated Legendre polynomials; and for $\alpha = -1/2$, to Chebyshev polynomials of the first kind. In a more general case, the parameter $\varepsilon$ controlling the attenuation coefficient is added (see Remark~\ref{remLegendre}). Thus, ultraspherical filters generalize the Legendre filters considered above, as well as Chebyshev filters of type I. Moreover, the limiting case ($\alpha \to +\infty$) for ultraspherical filters corresponds to Butterworth filters \cite{Paa_03}.

\section{Spectral Form of Mathematical Description (Fourier Basis)}\label{secSpectral}

As the basis $\BasisT$ of the space $\LPT$, we choose trigonometric functions:
\begin{equation}\label{eqDefFou}
  q(i,t) = \left\{ \begin{aligned}
    & \sqrt{1/T} & & \text{for} ~ i = 0 \\
    & \sqrt{2/T} \, \cos (i \pi t / T) & & \text{for} ~ i = 2k \\
    & \sqrt{2/T} \, \sin ((i+1) \pi t / T) & & \text{for} ~ i = 2k-1,
  \end{aligned} \right. \ \ \ \ \ \
  \begin{gathered}
    i = 0,1,2,\dots, \\
    k = 1,2,3,\dots
  \end{gathered}
\end{equation}

To find the two-dimensional nonstationary transfer function for any of the linear filters considered in this paper, we need the two-dimensional nonstationary transfer function $P$ of the differentiator or the two-dimensional nonstationary transfer function $P^{-1}$ of the integrator.

The elements of the matrix $P$ are defined as follows \cite{SolSemPeshNed_79}:
\[
  P_{ij} = 2 q(i,0) q(j,0) + \mathcal{P}_{ij},
\]
where
\[
  \mathcal{P}_{ij} = \int_\mathds{T} q(i,t) \, \frac{dq(j,t)}{dt} \, dt, \ \ \ i,j = 0,1,2,\dots,
\]
and the factor $2$ is introduced to reflect the property of the trigonometric Fourier series when approximating a discontinuous function (at the discontinuity point).

Applying integration by parts, we have
\[
  q(i,0) q(j,0) + \int_\mathds{T} q(i,t) \, \frac{dq(j,t)}{dt} \, dt = q(i,T) q(j,T) - \int_\mathds{T} q(j,t) \, \frac{dq(i,t)}{dt} \, dt;
\]
therefore, $\mathcal{P}_{ij} = -\mathcal{P}_{ji}$, since $q(i,0) = q(i,T)$ for the chosen basis, $i = 0,1,2,\dots$

For the trigonometric basis, the computation of $\mathcal{P}_{ij}$ is straightforward. The derivative of the basis function with index $j = 0$ is identically zero, while derivatives of the other basis functions are expressed in terms of basis functions with neighbouring indices. If $j = 2k$, then
\[
  \frac{dq(j,t)}{dt} = \sqrt{\frac{2}{T}} \biggl[ \cos \frac{j \pi t}{T} \biggr]' = -\sqrt{\frac{2}{T}} \, \frac{j \pi}{T} \, \sin \frac{j \pi t}{T} = -\frac{j \pi}{T} \, q(j-1,t),
\]
and if $j = 2k-1$, then
\[
  \frac{dq(j,t)}{dt} = \sqrt{\frac{2}{T}} \biggl[ \sin \frac{(j+1) \pi t}{T} \biggr]' = \sqrt{\frac{2}{T}} \, \frac{(j+1) \pi}{T} \, \cos \frac{(j+1) \pi t}{T} = \frac{(j+1) \pi}{T} \, q(j+1,t),
\]
which is consistent with $\mathcal{P}_{ij} = -\mathcal{P}_{ji}$.

Thus, the nonzero elements $\mathcal{P}_{ij}$ are
\begin{gather*}
  \mathcal{P}_{m-1,m} = -\mathcal{P}_{m,m-1} = - \frac{m\pi}{T}, \ \ \ m = 2,4,6,\dots
\end{gather*}

Taking into account the values of basis functions at $t = 0$:
\[
  q(i,0) = \left\{ \begin{aligned}
    & \sqrt{1/T} & & \text{for} ~ i = 0 \\
    & \sqrt{2/T} & & \text{for} ~ i = 2k \\
    & 0\vphantom{\sqrt{T}} & & \text{for} ~ i = 2k-1,
  \end{aligned} \right. \ \ \ \ \ \
  \begin{gathered}
    i = 0,1,2,\dots, \\
    k = 1,2,3,\dots,
  \end{gathered}
\]
we obtain the nonzero elements of the matrix $P$:
\begin{gather*}
  P_{00} = \frac{2}{T}, \ \ \ P_{0m} = P_{m0} = \frac{2\sqrt{2}}{T}, \ \ \ P_{ml} = \frac{4}{T}, \\
  P_{m-1,m} = -P_{m,m-1} = - \frac{m\pi}{T}, \ \ \ m,l = 2,4,6,\dots
\end{gather*}

Further, we give the formula for elements of $P^{-1}$ \cite{SolSemPeshNed_79}:
\[
  P_{ij}^{-1} = \int_\mathds{T} q(i,t) \int_0^t q(j,\tau) d\tau dt, \ \ \ i,j = 0,1,2,\dots
\]

Using integration by parts, we find
\[
  \int_\mathds{T} q(i,t) \int_0^t q(j,\tau) d\tau dt = \int_\mathds{T} q(i,t) dt \int_\mathds{T} q(j,\tau) d\tau - \int_\mathds{T} q(j,t) \int_0^t q(j,\tau) d\tau dt.
\]
Furthermore, since $q(0,t) = \sqrt{1/T}$, from the orthonormality of the basis we have
\[
  \int_\mathds{T} q(i,t) dt = \sqrt{T} \int_\mathds{T} q(0,t) q(i,t) dt = \sqrt{T} \delta_{0i},
\]
where $\delta_{0i}$ is the Kronecker delta. Hence,
\[
  P_{ij}^{-1} = T \delta_{0i} \delta_{0j} - P_{ji}^{-1}.
\]

This property implies that $P_{00}^{-1} = T/2$ and $P_{ii}^{-1} = 0$ for $i = 1,2,\dots$, and also $P^{-1}_{ij} = -P^{-1}_{ji}$ when $i^2 + j^2 \neq 0$. Moreover, it suffices to consider the case $i < j$. Clearly, the antiderivative of the basis function with index $j \neq 0$ is expressed in terms of basis functions with neighbouring indices and possibly through the basis function with index $j = 0$. If $j = 2k$, then
\[
  \int_0^t {q(j,\tau) d\tau} = \sqrt{\frac{2}{T}} \int_0^t {\cos \frac{j \pi \tau}{T} \, d\tau} = \frac{\sqrt{2T}}{j \pi} \sin \frac{j \pi t}{T} = \frac{T}{j \pi} \, q(j-1,t),
\]
and if $j = 2k-1$, then
\begin{align*}
  \int_0^t {q(j,\tau) d\tau} & = \sqrt{\frac{2}{T}} \int_0^t {\sin \frac{(j+1) \pi \tau}{T} \, d\tau} \\
  & = \frac{\sqrt{2T}}{(j+1) \pi} \biggl[ 1 - \cos \frac{(j+1) \pi t}{T} \biggr] = \frac{T}{(j+1) \pi} \biggl[ \sqrt{2} q(0,t) - q(j+1,t) \biggr].
\end{align*}

Thus, we obtain the nonzero elements of the matrix $P^{-1}$:
\begin{gather*}
  P_{00}^{-1} = \frac{T}{2}, \ \ \ P_{0m}^{-1} = -P_{m0}^{-1} = \frac{\sqrt{2} T}{(m+1) \pi}, \ \ \ P_{l-1,l}^{-1} = -P_{l,l-1}^{-1} = \frac{T}{l \pi}, \\
  k = 1,2,3,\dots, \ \ \ m = 2k-1, \ \ \ l = 2k.
\end{gather*}

As standard signals, we consider $f(t) = \sin \omega t$ and $f(t) = \cos \omega t$, where $\omega > 0$, and find the corresponding nonstationary spectral characteristics.

The formulae presented below are based on well-known trigonometric identities \cite{KornKorn_00}:
\begin{align*}
  \sin \alpha t \sin \beta t & = \frac{\cos(\alpha - \beta) t - \cos(\alpha + \beta) t}{2}, \\
  \cos \alpha t \cos \beta t & = \frac{\cos(\alpha - \beta) t + \cos(\alpha + \beta) t}{2}, \\
  \sin \alpha t \cos \beta t & = \frac{\sin(\alpha + \beta) t + \sin(\alpha - \beta) t}{2},
\end{align*}
so their detailed derivation is omitted.

This applies to the formulae for computing both nonstationary spectral characteristics of standard signals and two-dimensional nonstationary transfer functions of the two additional blocks.

The elements of the column matrix $F$ corresponding to $f(t) = \sin \omega t$ can be found as follows:
\begin{align*}
  & F_0 = \frac{1 - \cos T \omega}{\sqrt{T} \omega}; \\
  & \text{if} \ \ \ i = 2k, \ \ \ \text{then} \\
  & \ \ \ F_i = \frac{T \sqrt{2T} \omega [(-1)^i \cos T \omega - 1]}{i^2 \pi^2 - T^2 \omega^2} \ \ \ \text{for} \ \ \ i \pi \neq T \omega, \ \ \ F_i = 0 \ \ \ \text{for} \ \ \ i \pi = T \omega; \\
  & \text{if} \ \ \ i = 2k-1, \ \ \ \text{then} \\
  & \ \ \ F_i = -\frac{\pi \sqrt{2T} (i+1) \sin T \omega}{(i+1)^2 \pi^2 - T^2 \omega^2} \ \ \ \text{for} \ \ \ (i+1) \pi \neq T \omega, \ \ \ F_i = \sqrt{\frac{T}{2}} \ \ \ \text{for} \ \ \ (i+1) \pi = T \omega; \\
  & k = 1,2,3,\dots,
\end{align*}
and for the column matrix $F$ corresponding to $f(t) = \cos \omega t$, the respective expressions are
\begin{align*}
  & F_0 = \frac{\sin T \omega}{\sqrt{T} \omega}; \\
  & \text{if} \ \ \ i = 2k, \ \ \ \text{then} \\
  & \ \ \ F_i = -\frac{T \sqrt{2T} \omega \sin T \omega}{i^2 \pi^2 - T^2 \omega^2} \ \ \ \text{for} \ \ \ i \pi \neq T \omega, \ \ \ F_i = \sqrt{\frac{T}{2}} \ \ \ \text{for} \ \ \ i \pi = T \omega; \\
  & \text{if} \ \ \ i = 2k-1, \ \ \ \text{then} \\
  & \ \ \ F_i = \frac{\pi \sqrt{2T} (i+1) (1 - \cos T \omega)}{(i+1)^2 \pi^2 - T^2 \omega^2} \ \ \ \text{for} \ \ \ (i+1) \pi \neq T \omega, \ \ \ F_i = 0 \ \ \ \text{for} \ \ \ (i+1) \pi = T \omega; \\
  & k = 1,2,3,\dots
\end{align*}

Next, we consider the two-dimensional nonstationary transfer function $S^\tau$ of a pure time shift element by $\tau$, where $\tau < 0$ and $\tau > 0$ correspond to delay and advance, respectively. This accounts for phase delay in the spectral domain without returning to the time domain. The elements of $S^\tau$ are given by
\[
  S_{ij}^\tau = \int_\mathds{T} q(i,t) q(j,t+\tau) dt, \ \ \ i,j = 0,1,2,\dots,
\]
where the result depends on how the basis functions $q(i,t)$ are extended outside the interval $\mathds{T}$. We assume that the values of basis functions outside $\mathds{T}$ are computed by the same rule as on $\mathds{T}$. Trigonometric functions are periodic, so their behavior outside $\mathds{T}$ is the same as on the interval. This approach has proved effective for trigonometric functions \cite{RybShe_FraOp26} (in \cite{RybShe_FraOp26}, extension by zero was also considered for comparison), but it may not be applicable for orthogonal polynomials.

The nonzero elements of the matrix $S^\tau$ satisfy the following relations:
\begin{gather*}
  S_{00}^\tau = 1, \ \ \ S_{l-1,l-1}^\tau = S_{ll}^\tau = \cos \frac{l \pi \tau}{T}, \ \ \ S_{l-1,l}^\tau = -S_{l,l-1}^\tau = -\sin \frac{l \pi \tau}{T}, \\
  k = 1,2,3,\dots, \ \ \ l = 2k.
\end{gather*}

The matrix $S^\tau$ has a block-diagonal structure with $2 \times 2$ blocks that are rotation matrices by angles depending on $\tau$.

For error computation, we need the two-dimensional nonstationary transfer function $A^\theta$ of an amplifier with gain $a(t) = \chi_{[0,\theta]}(t)$, where $\chi_{[0,\theta]}(t)$ is the indicator of the set $[0,\theta]$, $0 < \theta \leqslant T$. Its elements can be computed according to the expression
\[
  A_{ij}^\theta = \int_\mathds{T} \chi_{[0,\theta]}(t) q(i,t) q(j,t) dt = \int_0^\theta q(i,t) q(j,t) dt, \ \ \ i,j = 0,1,2,\dots
\]

Its use is associated with accounting for phase delay and with the fact that the filtering problem is solved on a finite time interval.

As a result of computations, we find
\begin{align*}
  & A_{00}^\theta = \frac{\theta}{T}; \\
  & \text{if} \ \ \ i = 2k, \ \ \ \text{then} \\
  & \ \ \ A_{0i}^\theta = A_{i0}^\theta = \frac{\sqrt{2}}{i \pi} \sin \frac{i \pi \theta}{T}, \ \ \
  A_{ii}^\theta = \frac{\theta}{T} + \frac{1}{2i \pi} \sin \frac{2i \pi \theta}{T}, \ \ \ i > 0, \\
  & \ \ \ A_{i,i-1}^\theta = A_{i-1,i}^\theta = \frac{1}{2i \pi} \biggl( 1 - \cos \frac{2i \pi \theta}{T} \biggr), \ \ \ i > 1; \\
  & \text{if} \ \ \ i = 2k-1, \ \ \ \text{then} \\
  & \ \ \ A_{0i}^\theta = A_{i0}^\theta = \frac{\sqrt{2}}{(i+1) \pi} \biggl( 1 - \cos \frac{(i+1) \pi \theta}{T} \biggr), \ \ \
  A_{ii}^\theta = \frac{\theta}{T} - \frac{1}{2(i+1) \pi} \sin \frac{2(i+1) \pi \theta}{T}, \ \ \ i > 0, \\
  & \ \ \ A_{i,i-1}^\theta = A_{i-1,i}^\theta = \frac{1}{2i \pi} \biggl( 1 - \cos \frac{2i \pi \theta}{T} + i \biggl( 1 - \cos \frac{2 \pi \theta}{T} \biggr) \biggr), \ \ \ i > 1; \\
  & \text{if} \ \ \ i = 2k > 0 \ \ \ \text{and} \ \ \ j = 2l > 0, \ \ \ \text{then} \\
  & \ \ \ A_{ij}^\theta = A_{ji}^\theta = \frac{2}{(i^2 - j^2) \pi} \biggl( i \sin \frac{i \pi \theta}{T} \cos \frac{j \pi \theta}{T} - j \cos \frac{i \pi \theta}{T} \sin \frac{j \pi \theta}{T} \biggr); \\
  & \text{if} \ \ \ i = 2k-1 \ \ \ \text{and} \ \ \ j = 2l-1, \ \ \ \text{then} \\
  & \ \ \ A_{ij}^\theta = A_{ji}^\theta = \frac{2}{((i+1)^2 - (j+1)^2) \pi} \\
  & \ \ \ \ \ \ {} \times \biggl( (j+1) \sin \frac{(i+1) \pi \theta}{T} \cos \frac{(j+1) \pi \theta}{T} - (i+1) \cos \frac{(i+1) \pi \theta}{T} \sin \frac{(j+1) \pi \theta}{T} \biggr); \\
  & \text{if} \ \ \ i = 2k > 0 \ \ \ \text{and} \ \ \ j = 2l-1 > 0, \ \ \ \text{then} \\
  & \ \ \ A_{ij}^\theta = A_{ji}^\theta = \frac{2}{(i^2 - j^2) \pi} \biggl( i \sin \frac{i \pi \theta}{T} \sin \frac{(j+1) \pi \theta}{T} + (j+1) \biggl( \cos \frac{i \pi \theta}{T} \cos \frac{(j+1) \pi \theta}{T} - 1\biggr) \biggr); \\
  & k,l = 1,2,3,\dots
\end{align*}

Note that if $\tau = 0$, then $S^\tau = E$ (the pure time shift element with $\tau = 0$ is the identity operator), and if $\theta = T$, then $A^\theta = E$ (the amplifier with gain $a(t) \equiv 1$ is also the identity operator).

\section{Testing Methodology and Numerical Experiment}\label{secNumerical}

In this section, we use the testing methodology from \cite{RybShe_FraOp26}. The difference is that only deterministic noise is considered here, since this suffices to demonstrate the proposed technique (in \cite{RybShe_FraOp26}, not only deterministic but also random noise was considered). The example for the numerical experiment is similar to that in \cite{RybShe_FraOp26}, but the numerical parameters in this paper are chosen for a better fit to the Bessel filter and the trigonometric basis.

The spectral form of mathematical description implies that all operations with signals are performed in the spectral domain (filtering, phase delay compensation, error computation). Signals in the time domain are needed only for problem formulation and visualization of the output signals.

We consider the useful signal and noise:
\begin{gather*}
  u(t) = \sin 5 \pi t \ \ \ \text{and} \ \ \ v(t) = \sigma (\cos 85 \pi t + \sin 105 \pi t), \\
  \sigma = 0.1, \ \ \ t \in \mathds{T} = [0,1] \ \ \ (T = 1),
\end{gather*}
where the odd coefficients $5$, $85$, and $105$ are chosen so that the functions $u(t)$ and $v(t)$ have nontrivial nonstationary spectral characteristics (trivial spectral characteristics contain only one or several nonzero elements).

The filter input is the signal $g(t) = u(t) + v(t)$ (an additive mixture of the useful signal and noise); then the filter output $x(t)$ is compared to the useful signal $u(t)$.

We introduce the notation
\[
  U = \mathbb{S} [u], \ \ \ V = \mathbb{S} [v],
\]
then
\[
  G = \mathbb{S} [g] = U + V,
\]
where
\begin{gather*}
  V = \sigma (V_1 + V_2); \\
  \begin{aligned}
    V_1 & = \mathbb{S} [v_1], & v_1(t) & = \cos 85 \pi t; \\
    V_2 & = \mathbb{S} [v_2], & v_2(t) & = \sin 105 \pi t.
  \end{aligned}
\end{gather*}

All the necessary formulae for elements of the nonstationary spectral characteristics $U,V_1,V_2$ are presented in Section~\ref{secSpectral}.

The nonstationary spectral characteristic $X = \mathbb{S} [x]$ of the output signal $x(t)$ is given by expression~\eqref{eqInOutSp}, in which the two-dimensional nonstationary transfer function $W$ is determined by the choice of the filter (Bessel, Papoulis, or Legendre filters; see Section~\ref{secFilterExamples}). The two-dimensional nonstationary transfer function $W$ also depends on the cutoff frequency $\hat \Omega$ and additional parameters, if any (e.g., the parameter $m$ for Legendre filters). To return to the time domain, we use the inversion formula~\eqref{eqSpInvert}.

The filtering quality is conveniently expressed in terms of the error that shows the difference between the useful signal $u(t)$ and the output signal $x(t)$. Direct comparison is inefficient due to phase delay; therefore, when computing the error, two additional transformations are used.

The first transformation compensates for the phase delay~\cite{LutTosEva_01}, i.e., we shift the signal to $x^*(t) = x(t + \tau_\varphi)$, where
\[
  \tau_\varphi = -\frac{\arg H(\mi \omega)}{\omega} \bigg|_{\omega = \Omega} \ \ \ (\Omega = 5 \pi),
\]
and $H(\mi \omega)$ is the frequency response of the chosen filter. The second transformation restricts the domain, since comparison of signals on $[T-\tau_\varphi,T]$ is meaningless.

The corresponding formula for the error was proposed in \cite{RybShe_FraOp26}:
\[
  \mathcal{E} = \| x^* - u \|_\LPTs = \| \sqrt{\chi} \, (x^* - u) \|_\LPT,
\]
where $\chi_{[0,\theta]}(t)$ is the indicator of the set $\mathds{T}^* = [0,\theta]$ with $\theta = T - \tau_\varphi$, i.e.,
\[
  \mathcal{E}^2 = \int_0^{T-\tau_\varphi} \bigl( x^*(t) - u(t) \bigr)^2 dt = \int_0^T \chi_{[0,T-\tau_\varphi]}(t) \bigl( x^*(t) - u(t) \bigr)^2 dt.
\]

Its spectral counterpart is also given there:
\[
  \mathcal{E}^2 = (X^* - U)^\trans A^\theta (X^* - U) \ \ \ \text{and} \ \ \ \mathcal{E} = \sqrt{(X^* - U)^\trans A^\theta (X^* - U)},
\]
where $(\,\cdot\,)^\trans$ denotes transposition, and
\[
  X^* = S^{\tau_\varphi} W G.
\]

The transition to the time domain uses the inversion formula~\eqref{eqSpInvert}:
\[
  x(t) = \sum\limits_{i=0}^\infty X_i q(i,t) \ \ \ \text{and} \ \ \ x^*(t) = \sum\limits_{i=0}^\infty X_i^* q(i,t),
\]
where $X_i$ and $X_i^*$ are elements of the nonstationary spectral characteristics $X$ and $X^*$, respectively.

For comparison, we can use the a priori error, i.e., the norm of the function $v(t)$ in the space $\LPTs$:
\[
  \mathcal{E}_0 = \| v \|_\LPTs = \| \sqrt{\chi} \, v \|_\LPT,
\]
for which the spectral counterpart is also available \cite{RybShe_FraOp26}:
\[
  \mathcal{E}_0 = \sqrt{V^\trans A^\theta V}.
\]

However, the set $\mathds{T}^*$ is determined by the condition $\theta = T - \tau_\varphi$ and therefore depends on both the chosen filter family and the order of a particular filter. To avoid complicating the computations, we use the following estimate:
\[
  \mathcal{E}_0^+ = \| v \|_\LPT = \sqrt{V^\trans V},
\]
where $\mathcal{E}_0 \leqslant \mathcal{E}_0^+$, since $[0,\theta] \subseteq [0,T]$. The estimate $\mathcal{E}_0^+$ is convenient because it does not depend on the filter used.

For computations, all the infinite column matrices and infinite matrices must be truncated to finite sizes. Thus, nonstationary spectral characteristics become column matrices of size $L$, and two-dimensional nonstationary transfer functions, after truncation, have size $L \times L$. Then
\[
  x(t) \approx \sum\limits_{i=0}^{L-1} X_i q(i,t) \ \ \ \text{and} \ \ \ x^*(t) \approx \sum\limits_{i=0}^{L-1} X_i^* q(i,t).
\]

Truncation of nonstationary spectral characteristics and two-dimensional nonstationary transfer functions introduces an additional error; therefore, with truncation taken into account, the error $\mathcal{E}$ consists of two components. The first one is the filter error (the paper considers linear filters whose amplitude frequency responses approximate the amplitude frequency response of an ideal filter). The second one is the spectral method error associated with the truncation of all nonstationary spectral characteristics and two-dimensional nonstationary transfer functions. This component should decrease as the truncation order $L$ increases. The influence of both components can be observed by analyzing the computational results.

Below, we present the computational results for Bessel, Papoulis, and Legendre filters (two variants of Legendre filters are used). For each filter family, the errors $\mathcal{E}$ are given. For all the filters, the orders $n = 2,3,\dots,6$ are chosen. The truncation orders defining the matrix sizes are $L = 128, 256, 512, 1024$. The errors for Bessel filters are given in Table~\ref{tabBesselResult}, for Papoulis filters in Table~\ref{tabPapoulisResult}, and for Legendre filters with $m = 1$ and $m = 2$ in Tables~\ref{tabLegendre1Result} and~\ref{tabLegendre2Result}, respectively. The cutoff frequency is the same for all filters ($\hat \Omega = 25 \pi$).

The choice of the truncation order $L = 128$ ensures that the nonstationary spectral characteristic $V$ correctly accounts for the harmonic signals $\cos 85 \pi t$ and $\sin 105 \pi t$. This truncation order also enables a comparison of filtering quality if the Walsh and Haar bases are used instead of the trigonometric basis. The subsequent doubling of the truncation order ($L = 256, 512, 1024$) is convenient for analyzing the dependence of the error on $L$.

Next, we provide the exact value of the a priori error estimate: $\mathcal{E}_0^+ = \| v \|_\LPT = \sigma = 0.1$, and with truncation taken into account, $\mathcal{E}_0^+ = \sqrt{V^\trans V} = 0.099628, 0.099909, 0.099959, 0.099980$ for $L = 128, 256, 512, 1024$, respectively. Comparison of these values with the a priori errors (e.g., for the Bessel filter of order $n = 3$ we have $\mathcal{E}_0 = \sqrt{V^\trans A^\theta V} = 0.099443, 0.099619, 0.099619, 0.099666$) shows that for the analysis of the results, $\mathcal{E}_0$ can indeed be replaced by $\mathcal{E}_0^+$.

\ifnum \showtables = 1

\begin{table}[ht]
  \centering
  \renewcommand{\arraystretch}{1.1}

\caption{Error $\mathcal{E}$ for Bessel filters}\label{tabBesselResult}
\begin{tabular}{ccccc}
  \hline
  $~~n~~$ & $L = 128$ & $L = 256$ & $L = 512$ & $L = 1024$ \\
  \hline
  2 & 0.020384 & 0.020045 & 0.019954 & 0.019942 \\
  3 & 0.022963 & 0.022700 & 0.022626 & 0.022620 \\
  4 & 0.030673 & 0.030537 & 0.030493 & 0.030495 \\
  5 & 0.040459 & 0.040418 & 0.040395 & 0.040402 \\
  6 & 0.049117 & 0.049131 & 0.049119 & 0.049127 \\
  \hline
\end{tabular}

\vskip 2ex

\caption{Error $\mathcal{E}$ for Papoulis filters}\label{tabPapoulisResult}
\begin{tabular}{ccccc}
  \hline
  $~~n~~$ & $L = 128$ & $L = 256$ & $L = 512$ & $L = 1024$ \\
  \hline
  2 & 0.009558 & 0.008849 & 0.008570 & 0.008361 \\
  3 & 0.015965 & 0.014426 & 0.013685 & 0.013317 \\
  4 & 0.013371 & 0.010342 & 0.008566 & 0.007556 \\
  5 & 0.018551 & 0.015640 & 0.013853 & 0.012879 \\
  6 & 0.018983 & 0.014525 & 0.011885 & 0.010353 \\
  \hline
\end{tabular}

\vskip 2ex

\caption{Error $\mathcal{E}$ for Legendre filters ($m = 1$)}\label{tabLegendre1Result}
\begin{tabular}{ccccc}
  \hline
  $~~n~~$ & $L = 128$ & $L = 256$ & $L = 512$ & $L = 1024$ \\
  \hline
  2 & 0.012210 & 0.011478 & 0.011230 & 0.011074 \\
  3 & 0.012430 & 0.010429 & 0.009380 & 0.008893 \\
  4 & 0.014131 & 0.011457 & 0.009894 & 0.009045 \\
  5 & 0.016488 & 0.013000 & 0.010987 & 0.009773 \\
  6 & 0.019043 & 0.014743 & 0.012210 & 0.010773 \\
  \hline
\end{tabular}

\vskip 2ex

\caption{Error $\mathcal{E}$ for Legendre filters ($m = 2$)}\label{tabLegendre2Result}
\begin{tabular}{ccccc}
  \hline
  $~~n~~$ & $L = 128$ & $L = 256$ & $L = 512$ & $L = 1024$ \\
  \hline
  2 & 0.010433 & 0.009669 & 0.009391 & 0.009194 \\
  3 & 0.010843 & 0.008608 & 0.007463 & 0.006865 \\
  4 & 0.012900 & 0.010392 & 0.008719 & 0.007799 \\
  5 & 0.015944 & 0.012289 & 0.010190 & 0.008996 \\
  6 & 0.018383 & 0.014047 & 0.011623 & 0.010186 \\
  \hline
\end{tabular}
\end{table}

\fi

Each of the considered filters significantly suppresses the noise, since the errors $\mathcal{E}$ are smaller than the a priori error estimates $\mathcal{E}_0^+$. We now analyze the results in more detail for each filter family.

Bessel filters show the lowest accuracy. They are characterized by a slow roll-off of the amplitude frequency response \cite{Tho_PIEE49}; therefore, the harmonic signals $\cos 85 \pi t$ and $\sin 105 \pi t$ are suppressed less effectively than by filters of other types. Moreover, increasing $L$ does not noticeably affect the filtering quality. This indicates that for the error $\mathcal{E}$, the first component (the filter error) dominates, while the second one (the spectral method error) is negligible.

Papoulis filters yield a high-accuracy result; however, it depends on whether the filter order is even or odd. The accuracy of even-order filters is higher compared to that of odd-order filters (with one exception: the filter of order $n = 5$ is more accurate than the filter of order $n = 6$ for $L = 128$). This behavior is attributed to the specific features of their synthesis \cite{Pap_PIRE58, Pap_PIRE59}. The dependence of the error $\mathcal{E}$ on the truncation order $L$ is more pronounced here. Consequently, both error components affect the result. As the filter order increases, the influence of the first component decreases while that of the second one increases; therefore, for higher filter orders, we recommend that $L$ be increased.

Legendre filters outperform Papoulis filters of odd orders in accuracy (especially for $m = 2$). For them, the same conclusion can be drawn regarding the error $\mathcal{E}$: for low orders, the first component dominates (increasing $L$ improves accuracy, but only slightly); as the filter order increases, the second one begins to dominate. For Legendre filters, the condition $m = 2$ yields a smaller error compared to $m = 1$. In general, Legendre filters with $m = 2$ provide the best results.

Plots of the functions $u(t)$ (\textcolor{red}{red}) and $g(t)$ (\textcolor{blue}{blue}) are shown in Figure~\ref{picOriginalSignal}, and plots illustrating the filtering results for some values of $n$ and $L$ are presented in Figures~\ref{picBesselResult}--\ref{picLegendre2Result}: for $u(t)$ (\textcolor{red}{red}) and $x^*(t)$ (\textcolor{blue}{blue}). The differences between the plots of $g(t)$ and $x^*(t)$ are clearly visible for the Bessel filter on the entire interval $[0,T]$, while for the other filters they are observed only on the interval $[T-\tau_\varphi,T]$. The filter orders for which the plots are shown correspond to Examples 1, 2, and 3 from Section~\ref{secFilterExamples}.

\ifnum \showfigures = 1

\begin{figure}[ht]
  \centering
  \includegraphics[scale = 0.85]{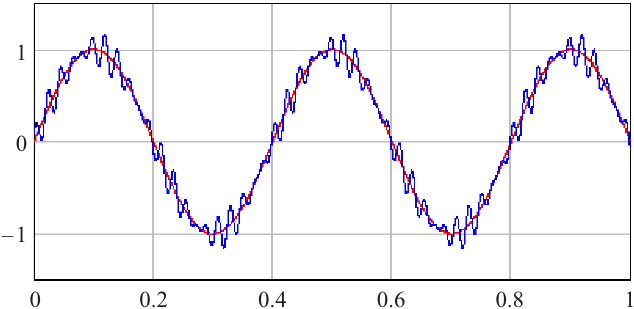}
  \caption{Plots of the useful signal and the sum of the useful signal and deterministic noise}\label{picOriginalSignal}
\end{figure}

\begin{figure}[p]
  \centering
  \includegraphics[scale = 0.85]{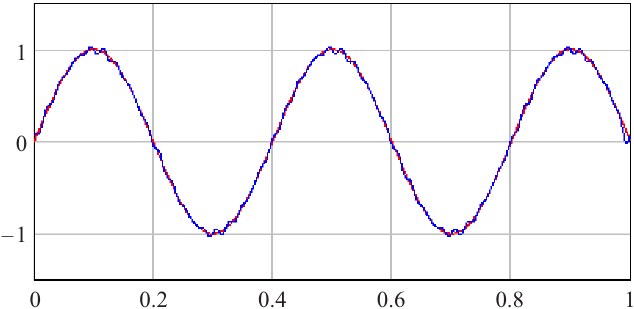}
  \caption{Plots of the useful signal and the output signal of the Bessel filter ($n = 3$, $L = 128$)}\label{picBesselResult}
  \vskip 2ex
  \includegraphics[scale = 0.85]{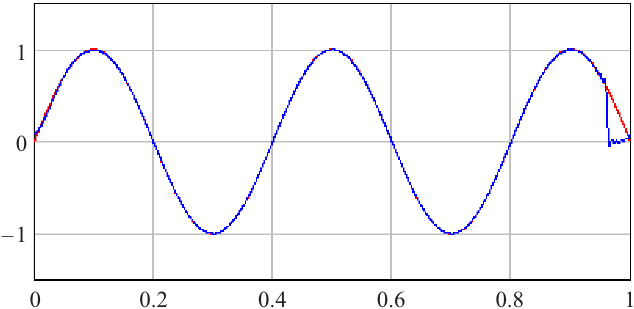}
  \caption{Plots of the useful signal and the output signal of the Papoulis filter ($n = 4$, $L = 256$)}\label{picPapoulisResult}
  \vskip 2ex
  \includegraphics[scale = 0.85]{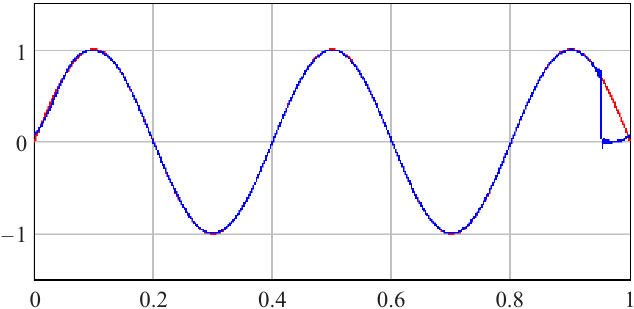}
  \caption{Plots of the useful signal and the output signal of the Legendre filter ($n = 5$, $m = 1$, $L = 512$)}\label{picLegendre1Result}
  \vskip 2ex
  \includegraphics[scale = 0.85]{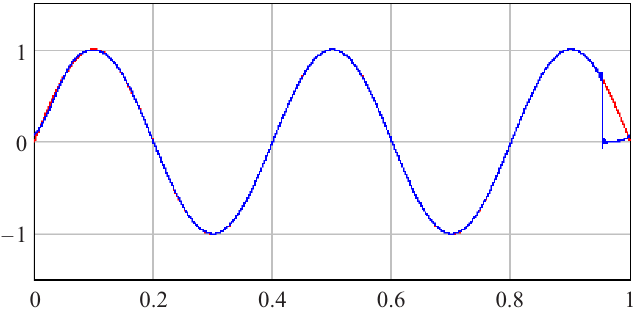}
  \caption{Plots of the useful signal and the output signal of the Legendre filter ($n = 5$, $m = 2$, $L = 1024$)}\label{picLegendre2Result}
\end{figure}

\fi

The discussion of the computational complexity of the proposed technique is given in \cite{RybShe_FraOp26}. It is based on well-known estimates of the complexity of matrix operations and therefore does not depend on the synthesis algorithm of a particular filter.

\section{Conclusions}\label{secSpConcl}

This paper describes a new technique for computer simulation of linear filters. Bessel, Papoulis, and Legendre filters of various orders are considered in detail. It also shows (more briefly) how the developed technique can be applied to modified Bessel filters, Halpern filters, and ultraspherical filters. The proposed technique is specifically intended for computer simulation; however, in contrast to many methods used by researchers in this field, it does not require a transition to discrete time, i.e., the filter output signal is modeled in continuous time. The results are based on the spectral form of mathematical description of linear control systems (the spectral method). They complement the previously obtained results for Butterworth, Linkwitz--Riley, and Chebyshev filters.

\end{document}